\documentclass[useAMS,usenatbib]{mn2e}

\usepackage{graphicx}

\title[The optical light curve of the LMC pulsar B0540-69]{The optical light curve of the LMC pulsar B0540-69 in 2009}

\author[S. Gradari et al.]
	{S. ~Gradari$^{1,5}$, M. ~Barbieri$^{1}$, C. ~Barbieri$^{1}$, G.~Naletto$^{2,3}$, E. ~Verroi$^{1,2}$,
	\newauthor
	T. ~Occhipinti$^{1,2}$, P. ~Zoccarato$^{1,5}$, C. ~German\`a$^{1}$, L. ~Zampieri$^{1}$,
	\newauthor
	A. ~Possenti$^{4}$\\
$^1$Department of Astronomy, University of Padova, Italy\\
$^2$Department of Information Engineering, University of Padova, Italy\\
$^3$CNR-IFN UOS Padova LUXOR, Padova, Italy\\
$^4$INAF - Astrophysical Observatory of Cagliari, Italy\\
$^5$CISAS, University of Padova, Italy}

\begin{document}

\date{accepted November 2010}

\pagerange{\pageref{firstpage}--\pageref{lastpage}} \pubyear{2010}

\maketitle

\label{firstpage}

\begin{abstract}

This paper reports a detailed analysis of the optical light curve of PSR B0540-69, the second brightest pulsar in the visible band, obtained in 2009 (Jan. 18 and 20, and Dec. 14, 15, 16, 18) with the very high speed photon counting photometer Iqueye mounted at the ESO 3.6-m NTT in La Silla (Chile). The optical light curve derived by Iqueye shows a double structure in the main peak, with a raising edge steeper than the trailing edge. The double peak can be fitted by two Gaussians with the same height and FWHM of 13.3 and 15.5 ms respectively. Our new values of spin frequencies allow to extend by 3.5 years the time interval over which a reliable estimate of frequency first and second derivatives can be performed. A discussion of implications on the braking index and age of the pulsar is carried out. A value of $n$ = 2.087 $\pm$ 0.007 for the overall braking index from 1987 to 2009 is derived. The braking index corrected age is confirmed around 1700 years.
\end{abstract}


\section{Introduction}

B0540-69 is a 50 ms optical pulsar in the Large Magellanic Cloud, the second brightest in the optical band after the Crab pulsar. It has been observed in recent years with a variety of imaging and spectroscopic instruments on ground as well as space telescopes (all references \cite{mign10}, \cite{delu07}, \cite{sera04} are based on HST data).  However, the number of published optical light curves is fairly small, and all amount to data obtained more than 10 years ago. The first published light curve was obtained by \cite{midd85} and \cite{midd87} using the 4-m and 1.5-m telescopes at Cerro Tololo. Then, \cite{goui92} derived a second light curve from data taken with the ESO 3.6-m telescope in the interval 1989 January Ð 1991 April. \cite{boyd95} obtained a third light curve with the High Speed Photometer (HSP) on board the HST. All light within the HSP sensitivity band from 160 to 700 nm was used, observing the pulsar for one hour on August 26, 1993 through a $0^{\prime\prime}.65$ diaphragm. The sample time was 300 microseconds. The HSP light curve was consistent with the shape seen by \cite{midd87} and \cite{goui92}, and showed with clarity a double peak structure. The HSP light curve was compared to those obtained by several hard X-ray instruments by \cite{depl03}, who fitted the pulse shape with a double Gaussian curve. A fourth curve, from data obtained in May 1994 at the ESO 3.6-m telescope, was inserted by \cite{mign98} in a paper about the pulsar PSR B1509-58. Subsequently, \cite{deet99} published a light curve based on data originally taken by \cite{manc89} at the 4-m Anglo Australian Telescope over the period 1986 14 July to 1988 16 June. Finally, \cite{ulme99} published a light curve obtained at CTIO in Nov. 1996, and suspected a strong phase difference between optical and X-ray data.
All those optical pulse shapes are consistent with the radio one \citep{manc93} and with those observed in the X- and Gamma-ray domains (e.g. \cite{mine99, deet99, depl03, camp08}).

We observed the pulsar at the ESO NTT with our very-high-speed photometer Iqueye (the prototype of a $^{\prime}$quantum$^{\prime}$ photometer for the European Extremely Large Telescope E-ELT, \cite{drav05}) in January and December 2009. Our optical light curve is therefore the first derived from data acquired since 1996. In Section 2 we describe the instrument and observations. The results from these observations are expounded in Section 3, and their implications on braking index and age are discussed in Section 4.

\section[]{The Iqueye Single Photon Counting Photometer}

For  a description of the scientific motivations of Iqueye and of its optomechanical characteristics see \cite{barb09} and \cite{nale09}, and references therein. In essence, Iqueye has been conceived as a precursor to a quantum photometer for the future 42-m European Extremely Large Telescope E-ELT (see \cite{drav05}), aiming to perform not only high speed photometry, but also the measurements of the statistics in the photon arrival times and intensity interferometry \citep{hanb74}. Iqueye is a conceptually simple fixed-aperture photometer which collects the light within a field of view (FOV) of few arcseconds around the target object. It is mounted at the Nasmyth focus of the 3.6-m ESO NTT.
A holed folding mirror at $45^{\circ}$ on the NTT focal plane brings a 1 arcmin field around the star under investigation to the field acquisition TV camera. The light from the target object instead passes through the central hole and is collected by a collimating refracting system. Two filter wheels located in the parallel beam after the first lens allow the selection of different filters and polarizers. Then the light reaches a focusing system which (de)magnifies the telescope image by a 1/3.25 factor. On this intermediate focal plane, one out of three pinholes (200, 300 and 500 micrometers diameter) can be inserted. These pinholes act as field stops, and their sizes allow the selection of three different FOVs (3.5, 5.2 and 8.7 arcsec diameter). After the pinhole, the light impinges on a pyramid having four reflecting surfaces and whose tip coincides with the center of the shadow of the secondary mirror. The pyramid splits the telescope pupil in four equal portions, and sends the light from each sub-pupil along four perpendicular arms. Along each arm, the sub-pupil light is first collimated and then refocused by a suitable system, further (de)magnifying the image by an additional 1/3.5 factor. Each sub-beam is then focused on a single photon avalanche photodiode (SPAD) operated in Geiger mode. The quantum efficiency of the Silicon SPADs extends from the blue to the near infrared, with a peak sensitivity of 55\% at 550 nm. When used without filter, as in the present case, the overall efficiency of Iqueye (SPAD + telescope + atmosphere) at the Zenith is approximately 33\%. The dark counts of the temperature-controlled detectors are very low, varying from 30 to 50 count/s for each individual unit. The SPAD circular sensitive area of 100 $\mu$m diameter, nominally defines a 5.8 arcsec FOV. Therefore, the smallest pinhole acts as the actual field stop at 3.4 arcsec. This pinhole can be selected when it is necessary to reduce as much as possible the background around the target, e.g. when observing a pulsar embedded in a nebula, as in the present case.
The optical solution of splitting the beam by a pyramid in 4 sub-beams was dictated by the need to overcome as much as possible the dead time intrinsic to the SPAD (75 nanoseconds), in order to give to Iqueye the largest possible dynamic range. In other scientific applications (e.g. intensity interferometry), having four independent detectors allows to cross correlate the counts from each sub-aperture.
The pulses produced by the SPADs, which have an intrinsic time jitter of the order of 35 ps, are sent to a Time to Digital Converter (TDC) board which has a nominal resolution of 24.4 ps. Considering also the other possible noise sources, the nominal accuracy of the photon arrival time determination is of the order of 100 ps or better. An external Rubidium oscillator provides the reference frequency to the TDC board. The board acquires also a pulse per second (PPS) from a GPS receiver, used to remove the Rubidium frequency drift and to put the internal detection times on the UTC scale. Taking into account all error factors, the final overall precision of each time tag in the UTC scale is approximately 450 ps, maintained throughout the duration of the observations. In order to take care as well as feasible of the rotation of the NTT building, the GPS antenna was mounted on the top of the dome, at the centre of one of the sliding doors (about 3 meters away from the dome rotation axis). The signal was brought to the receiver by a high-quality, length compensated cable. The geodetic and geocentric Cartesian coordinates of the antenna, in the WGS94 reference system, are given in Table \ref{coordinates}. These coordinates have been translated to the intersection of the optical and elevation axes using the construction drawings of the enclosure; taking into account the rotation of the dome, the actual precision is estimated better than 2 meters. This uncertainty on the position of the detector reflects in a maximum uncertainty on the barycentred times of arrival of the photons of 7 ns but does not influence significantly neither the values of the period nor of the phases, as we have also tested by changing the position of values by several tens of meters.

\begin{table*}
\caption{Geodetic and geocentric Cartesian coordinates of the NTT  (top of the roof, doors open, dome still)}
\begin{tabular}{@{}llrrrrlrlr@{}}
\hline
Long ($^{\circ}$)	&	Lat ($^{\circ}$)	&	Elevation (m)	&	X (m)	&	Y (m)	&	Z (m)\\

-70.733746	&	-29.258913	&	2424.09	&	+1838193.1	&	-5258983.2	&	-3100153.8\\
\hline
\end{tabular}
\label{coordinates}
\end{table*}

The user interface, developed as a Java multitasking code, controls each subsystem (e.g. the mechanisms), performs the data acquisition and storage, provides some real time monitoring of the data acquisition, and provides tools for a quick look statistical analysis of the data. Each arrival time is recorded on the storage device which has a total capacity of approximately 2 TB. Being the data stored in a mass memory device, all the data can be analyzed in post-processing: this allows, for example, to sort the collected time tags in arbitrarily long time bins still preserving the original data.
Between the January and December run, some improvements were made to the instrument \citep{barb10, nale10}, in particular the addition of a fifth SPAD to acquire the signal from the sky.

\section[]{Observations and analysis}

The observations were obtained through 3.5 or 5.2 arcsec diaphragms, without filters (maximum sensitivity around 550 nm, bandwidth at half maximum approximately 300 nm).
The observation log is provided in Table \ref{log}. The columns UTC and MJD = Modified Julian Date = JD - 2\,400\,000.5 provide values of time and date at mid counting period referred to the barycentre of the solar system in TCB units (see later for the adopted procedure).

\begin{table*}
\caption{Log of the observations of Iqueye at the NTT}
\begin{tabular}{@{}llrrrrlrlr@{}}
\hline
Date 		& UTC 			& MJD (d)					& Observation 	& Diaphragm	& Detected\\
			& (hh mm ss) 		& (mid-exposure 			& duration 	& diameter	& photons\\
			& 				& time)					& (s) 		& (arcsec)		\\
\hline
2009 01 18 	& 05 11 10.0 		& 54849.21665 				& 5994 		& 3.5		& 8 304 630\\
\hline
2009 01 20 	& 04 03 19.0 		& 54851.16190 				& 5874 		& 5.2		& 17 775 124\\
\hline
2009 12 14	& 07 27 59.9		& 55179.31111				& 3600		& 3.5		& 21 917 637\\
\hline
2009 12 15	& 02 42 00.0		& 55180.11250				& 3600		& 3.5		& 10 767 547\\
\hline
2009 12 16	& 01 39 59.6		& 55181.06944				& 3000		& 5.2		& 13 896 153\\
\hline
2009 12 18	& 02 30 00.3		& 55183.10417				& 3600		& 3.5		& 10 262 179\\
\hline
\end{tabular}
\label{log}
\end{table*}

The procedure we routinely follow at the telescope to centre faint pulsating objects is to bin the arrival times in convenient time bins, e.g. 1/20 of the expected period, so that standard time-series analysis algorithms can be applied to single out the frequencies in the signal. After few minutes of trying in a given position, a slightly different one is tested until the position giving the best signal is found. Then, a long observation is started. In the case of B0540-69, the procedure converged very quickly. The Power Spectral Density of the data was dominated by a frequency at the expected (according to the ephemerides available in the literature) value of 19.7433 Hz (period around 0.05065 s) for January's data and 19.7380 Hz (period around 0.05066 s) for December's data with a statistical significance higher than 20 standard deviations ($\sigma^{\prime}$s) of noise; no other signal was visible above 3$\sigma^{\prime}$s of noise in the range 0-200 Hz. In December, this standard procedure was greatly helped by the availability of a very deep finding chart, kindly provided by \cite{mign10} (their fig. 1) before publication.

In order to perform the detailed analysis of the period and light curve, the arrival times of the photons are referred to the barycentre of the solar system, by using the latest release of the Tempo2 software \citep{hobb06} with the DE405 JPL Ephemerides \citep{stan98}.
The assumed celestial coordinates of the source are RA2000 = 05h40m11s.202 $\pm$ 0s.009; DEC2000 = $-69^{\circ}19^{\prime}54^{\prime\prime}.17 \pm 0^{\prime\prime}.05$ \citep{mign10}, with zero proper motion (\cite{mign10, delu07}). These values have been measured on images taken with the HST/WFPC2 through the F555 and F547nm filters, and have been referred to the absolute reference frame using astrometric data of the dense Two Micron All Sky Survey (2MASS) catalogue \citep{skru06}. Therefore, the positions of \cite{mign10} are more accurate than those published by \cite{sera04} and by \cite{shea94}.

We can consider the width of the peak in the signal Fourier transform, $P^2/(2\Delta t)$, where $P$ is the period and $\Delta t$ the observing time, which corresponds to a few microseconds in our cases, as the intrinsic resolution of the period. So, we took advantage of the actual much better timing performance of the instrument to determine the period with the highest possible accuracy. For this, we analyzed a 3 $\mu$s wide Òintrinsic resolutionÓ window around the expected period in steps of 0.1-0.3 ns by means of  an epoch-folding technique similar to that expounded by \cite{leah83}. The spin period P was computed separately for each night, starting from the values given by \cite{livi05} and reported to our dates including the effect of the first and second spin period derivates. For each period, the $\chi^2$ values against the zero hypothesis of a flat curve was calculated, obtaining a well defined distribution peaked around the expected value. The best period was then obtained through a least squares fit of the $\chi^2$ distribution with a Gaussian curve. The mean value of the Gaussian is the best fitting period. The error on this measurement was calculated starting from the Monte Carlo estimate of \cite{leah87} for sinusoidal signals and correcting it following the approach of \cite{lars96}, that makes optimum usage of the full pulse shape information, or equivalently all its Fourier components. By comparing equations (4) and (5) of \cite{lars96} we see that the period error for a non sinusoidal signal is $\sigma_P=\sigma_{P,s}/\sqrt{\sum_k k^2 A_k^2/A^2}$, where $A_k$ are the amplitude of the different Fourier components and $\sigma_{P,s}$ is the period error if the signal were a sinusoid with the same period and with amplitude $A$. The accuracy of the period determination is increased taking a weighted mean of the harmonics of the signal. $\sigma_{P,s}$ can be expressed using equation (6a) for sinusoidal signals in \cite{leah87}, while the amplitudes $A_k$ are determined from a Fourier component analysis of the signal. $A$ is taken approximately equal to $A_1$. In Table \ref{period} we report the results of the procedure, in terms of both the measured period $P$ and the corresponding pulsar spin frequency $\nu$. The quoted errors on $P$ are determined as explained above, while those on $\nu$ are obtained by error propagation.

\begin{table*}
\caption{Periods and frequencies of PSR B0540-69 determined with Iqueye data obtained during 2009}
\begin{tabular}{@{}llrrrrlrlr@{}}
\hline
Date (MJD)	& Period (s)				& Frequency (Hz)\\
			& and error (s)				& and error (Hz)\\
\hline
54849.21665	& 0.050 649 974 5 (43.8$\times10^{-9}$)		& 19.743 346 6 (17.1$\times10^{-6}$)\\
\hline
54851.16190	& 0.050 650 017 3 (23.8$\times10^{-9}$)		& 19.743 329 9 (9.3$\times10^{-6}$)\\
\hline
55179.31111	& 0.050 663 549 8 (44.4$\times10^{-9}$)		& 19.738 056 3 (17.3$\times10^{-6}$)\\
\hline
55180.11250	& 0.050 663 632 9 (25.4$\times10^{-9}$)		& 19.738 023 9 (9.9$\times10^{-6}$)\\
\hline
55181.06944	& 0.050 663 671 5 (49.1$\times10^{-9}$)		& 19.738 008 9 (19.1$\times10^{-6}$)\\
\hline
55183.10417	& 0.050 663 753 2 (24.3$\times10^{-9}$)		& 19.737 977 1 (9.5$\times10^{-6}$)\\
\hline
\end{tabular}
\label{period}
\end{table*}

With our data alone, obtained over a time span of one year, we can determine the first derivative of the frequency. Using $t_0$ = 55183.1042 (MJD) as reference date in a linear fit, we obtain $\nu_0 = 19.7379712 \pm 4.83\times10^{-6}$ Hz and $\dot{\nu_0} = -1.86346\times10^{-10} \pm 2.65\times10^{-13}$ Hz/s. These values are in good agreement with those available in the literature as discussed at the end of the Section 4.

\begin{figure}
\includegraphics[scale=0.37]{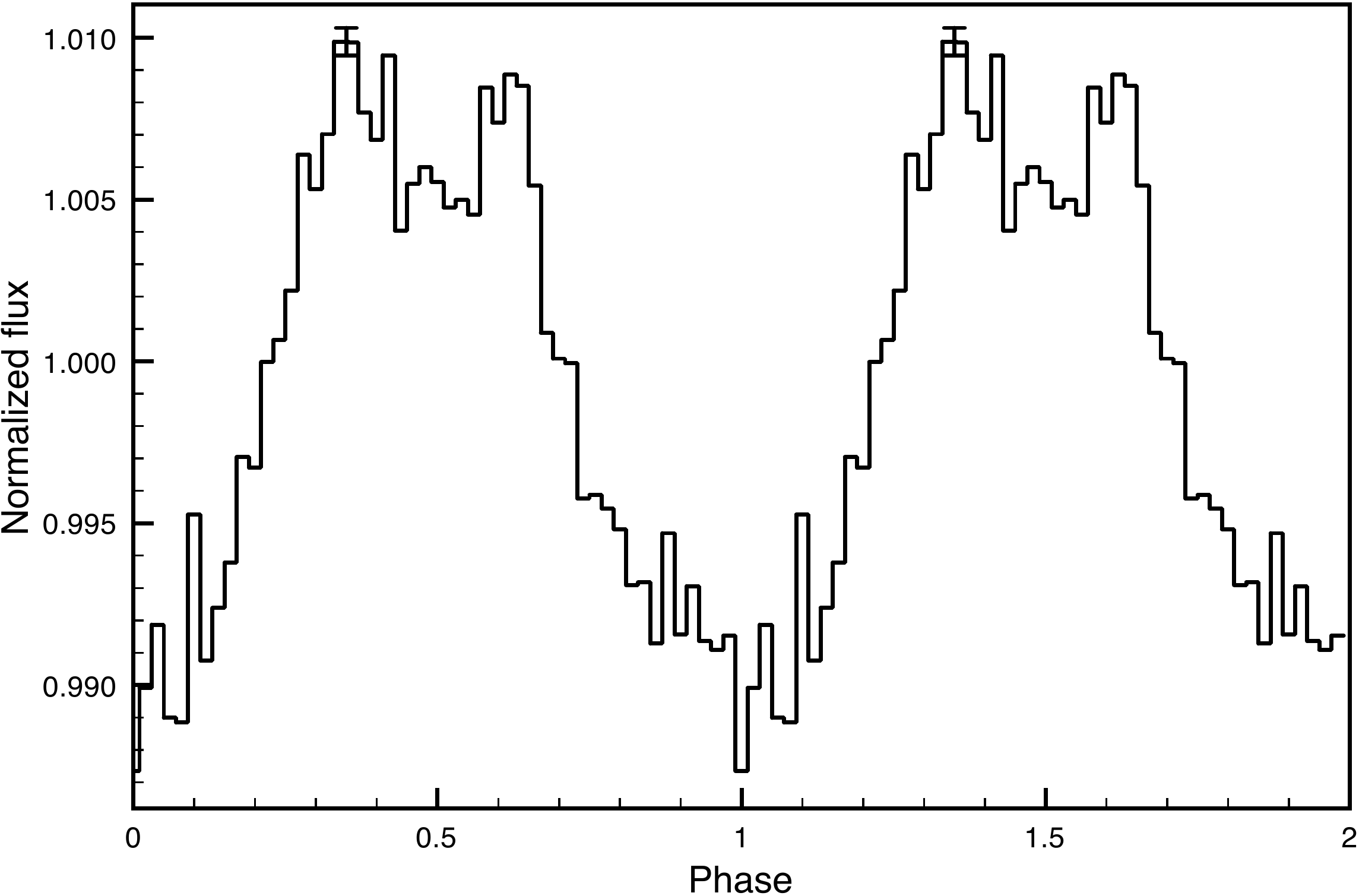}
\caption{The overall Iqueye light curve in January and December 2009, from the individual curves weighted according to the respective $\chi^2$ value and binned in 50 phase intervals. The counts have been normalized to the average count value during a period. For clarity the curve is shown over two cycles. The vertical bar shows the 1 sigma error. All light curves were shifted to match the phase of December 18.}
\label{combined}
\end{figure}

The combined Iqueye light curve for all nights of January and December 2009 is shown in Figure \ref{combined} using 50 phase bins and displayed for better visualization over two cycles. All light curves were shifted to match the phase of December 18 because of the better S/N ratio of those data. Each curve has been weighted for the respective $\chi^2$ value of the period determination and the alignment by minimizing the distance between the curves using again the $\chi^2$ method.

The modulation $M$ of our light curve, defined as
\begin{equation}
	M = \frac{<c>-c_{min}}{<c>} ,
\end{equation}
where $<c>$ is the mean count rate in the pulse profile and c$_{min}$ is the minimum count rate, is of the order of 1.5\% for the light curve obtained with Iqueye, a value similar to that shown by the light curves resulting from the published observations performed at other ground based telescopes.
Therefore, all available light curves published over the last 27 years have approximately the same modulation, and broadly show the same features, namely a main pulse with a total duty cycle of about 45\% and complex structure. Given the high number of acquired (indicatively, a mean rate of 2500 counts/sec) photons, and the extremely accurate time tagging guaranteed by Iqueye, we feel confident to say that the Iqueye light curve shown in Fig. \ref{combined} is the best available so far in visible light. The total duration of the main peak is approximately 22 ms (FWHM), with a central shallower feature suggesting the superposition of at least two peaks, as indicated also by the steeper slope of the ascending branch with respect to the descending one (0.062 $\pm$ 0.004 vs -0.046 $\pm$ 0.005 in units of normalized flux per phase unit) and already found by \cite{depl03} from X-ray data. Therefore we have fitted the broad central peak with two Gaussian components separated by 0.29 ($\pm$ 0.02) in phase, the leading one approximately 1.02 times higher than the second one, and with a FWHM of 13.3 $\pm$ 0.2 ms and 15.5 $\pm$ 0.2 ms respectively.

\section[]{Discussion of the braking index and age}

\begin{figure}
\includegraphics[scale=0.34]{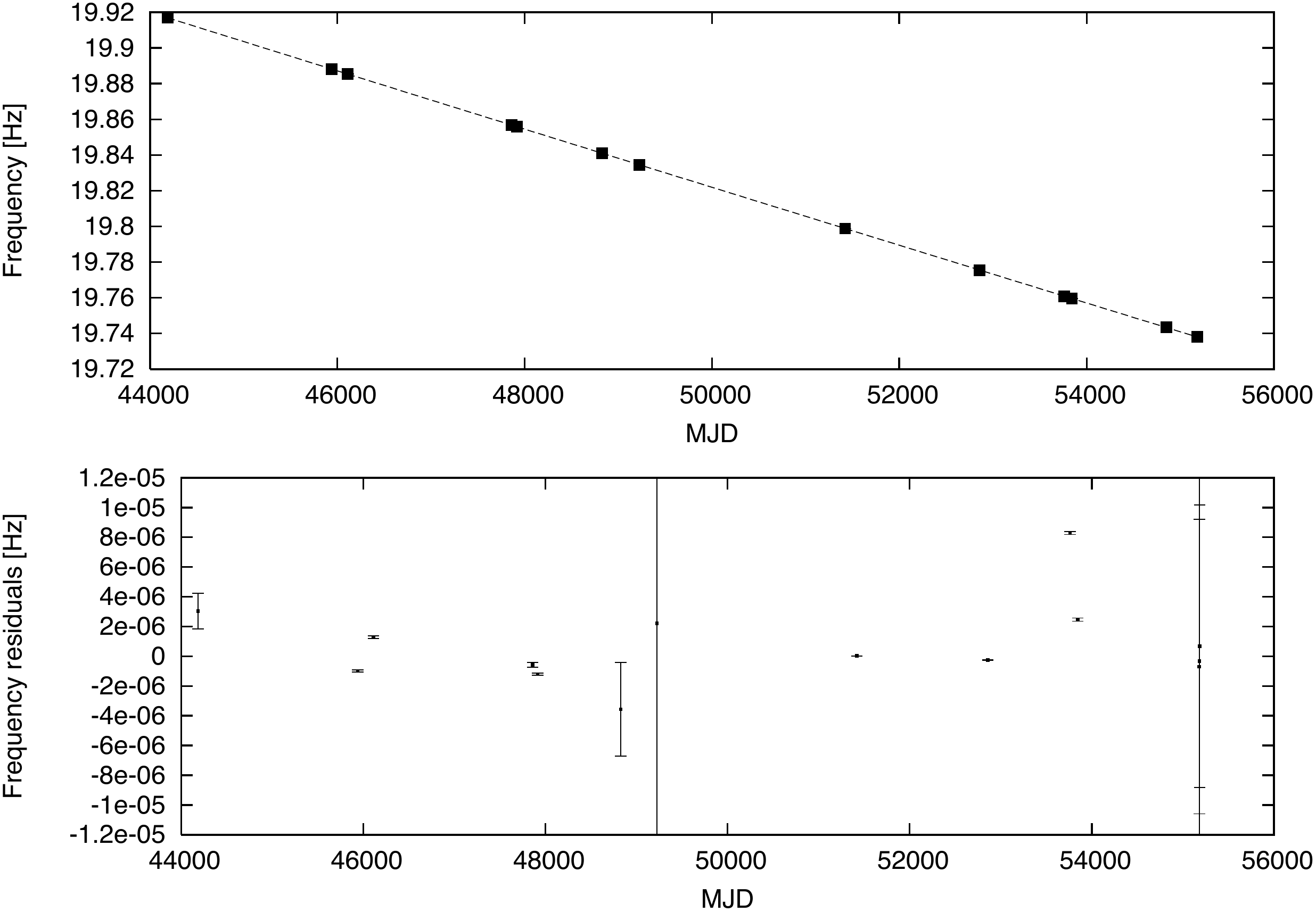}
\caption{Fit over the data showed in Table \ref{literature}}
\label{polynomial}
\end{figure}

In commonly assumed models for pulsar spin-down, a braking index $n$ and a characteristic age $\tau_c$ are defined \citep{livi07} by:
\begin{equation}
	\dot{\nu} = -K\nu^n ,\\
	n = \frac{\nu \ddot{\nu}}{\dot{\nu}^2} ,\\
	\tau_c = \frac{\nu}{2\dot{\nu}} ,
\end{equation}
where $\nu$ is the pulse frequency, $\dot{\nu}$ and $\ddot{\nu}$ are the first and second frequency derivatives respectively, and $K$ is a constant.
The braking index provides insight into the physics of the pulsar mechanism. Indeed, the actual value of the braking index is strictly related with the pulsar spin-down mechanism. It is well known \citep{manc77} that for magnetic dipole emission, as well as for the aligned rotator model of \citet{gold69}, $n$ = 3. Different values of $n$ would correspond to different processes of rotational energy loss and, in particular, values lower than 3 indicate that an additional torque is contributing to the spin-down. It should be noted that the value of $n$ also affects the determination of the pulsar age, with $n <$ 3 providing consistently larger values than those predicted by the characteristic age.
Among the additional spin-down mechanisms, the distortion of the magnetic dipole geometry, a time variable magnetic field, a change with time of the inclination angle between rotation and magnetic axes, and/or the presence of particles/currents in the magnetosphere have been suggested \citep{ghos07, livi07}. As for the Crab pulsar, in the case of PSR B0540-69 the existence of a synchrotron-emitting nebula around the pulsar provides independent evidence for a particle wind originating from the pulsar magnetosphere, whose plasma may then contribute an additional spin-down torque (as suggested also by \citet{boyd95}). However, precisely determining what physical mechanism is responsible for the observed braking index of PSR B0540-69 and pulsars in general remains a completely open question.
Unfortunately, the measurement of the braking index $n$ is a difficult task, and only the youngest pulsars (typical ages less than 2 kyears) possess all needed qualities, in particular rapid spin-down and small relatively infrequent glitches. As an example, a braking index $n$ = 2.51 $\pm$ 0.01 has been measured for the Crab pulsar. PSR B0540-69 bears many similarities to the Crab pulsar, like period- and magnetic- field strength (50 vs 30 ms, 5$\times$10$^{12}$ G  vs 4$\times$10$^{12}$ G, see \citet{camp08}), but its much larger distance prevents regular radio observations, and conflicting values of $n$ are reported in the literature. \citet{livi05} concluded, by a careful analysis of all X-ray data obtained using 7.6 years of data from the Rossi X-Ray Timing Explorer, that the best estimate for the braking index is $n$ = 2.140 $\pm$ 0.009. The optical data available until our observations provided the following values: \cite{manc89}, $n$ = 2.01 $\pm$ 0.02, \cite{goui92}, $n$ = 2.04 $\pm$ 0.02, and \cite{boyd95}, $n$ = 2.28 $\pm$ 0.02.
We calculated the first and second frequency derivatives adding the frequency values measured with Iqueye in January and December 2009 to the previously published data sets. The frequency values considered for our analysis are summarized in Table \ref{literature}. We have taken into account only measured, i.e. not interpolated, values, covering the entire spectrum from radio to X-ray at different dates. These values have been fitted with a second order polynomial (Fig. \ref{polynomial}), using least-squares regression, in the assumption of none or very small and infrequent glitches:

\begin{equation}
	\nu(t) = \nu(t_0) + \dot{\nu}(t-t_0) + \frac{1}{2}\ddot{\nu}(t-t_0)^2 .
\end{equation}

\begin{table*}
	\caption{Frequencies used for the calculation of the braking index. Values are taken from the corresponding papers indicated in the last column and ordered by MJD}
		\begin{tabular}{@{}llrrrrlrlr@{}}
		\hline
			{\bf MJD} 		& Frequency (Hz) 	& 	Band 		& Ref.\\
		\hline
			44186.91740 	& 19.91687532 		&	X 			& \cite{sewa84}\\
			45940.86590 	& 19.88811520 		&	VIS			& \cite{midd85}\\
			46111.07682 	& 19.88533133 		&	VIS			& \cite{midd85}\\
			47860.0000	& 19.85674751		&	Radio		& \cite{manc93}\\
			47915.0000	& 19.85584939		&	Radio		& \cite{manc93}\\
			48825.8000	& 19.84099663		&	Radio		& \cite{manc93}\\
			49225.25570 	& 19.83449650 		&	HST (UV+VIS) & \cite{boyd95}\\
			51421.62400 	& 19.79880010 		&	X 			& \cite{kaar01}\\
			52857.86600 	& 19.77553000 		&	X 			& \cite{john04}\\
			53761.76200 	& 19.76092260 		&	X 			& \cite{camp08}\\
			53843.56100 	& 19.75959520 		&	X 			& \cite{camp08}\\
			54849.21665 	& 19.74334657 		& 	VIS (Iqueye) 	& This work\\
			54851.16190 	& 19.74332988 		&	VIS (Iqueye) 	& This work\\
			55179.31111	& 19.73805633 		& 	VIS (Iqueye) 	& This work\\
			55180.11250	& 19.73802395 		& 	VIS (Iqueye) 	& This work\\
			55181.06944	& 19.73800890 		& 	VIS (Iqueye) 	& This work\\
			55183.10417	& 19.73797709 		& 	VIS (Iqueye) 	& This work\\
		\hline
		\end{tabular}
	\label{literature}
\end{table*}

The coefficients of the best fitting parabola, where $t_0$ is the value of the last date of observations, are reported in Table \ref{fit}.

\begin{table*}
	\caption{The coefficients of the second order polynomial used for the fit}
		\begin{tabular}{@{}llrrrrlrlr@{}}
		\hline
			 					& Value 					& Error\\
		\hline
			t$_0$ (MJD) 			& 55183.1042 				& \\
			$\nu_0$ (Hz) 			& 19.7379764 				& $\pm 1\times10^{-6}$\\
			$\dot{\nu_0}$ (Hz/s) 		& -1.86560$\times$10$^{-10}$	& $\pm 5\times10^{-15}$\\
			$\ddot{\nu_0}$ (Hz/s$^2$)	& 3.66$\times$10$^{-21}$ 	& $\pm 1\times10^{-23}$\\
		\hline
		\end{tabular}
	\label{fit}
\end{table*}

With the so determined values of $t_0$, $\nu_0$, $\dot{\nu_0}$ and $\ddot{\nu_0}$, the resulting value for the braking index is $n$ = 2.087 $\pm$ 0.007, and the characteristic age is $\tau$ = 1677.5 years. Our result is consistent within 3 combined sigmas with the value given by \cite{livi05} and it confirms that the value of the braking index for this pulsar is definitely lower than 3.

\section{Conclusions}

We have observed the LMC B0540-69 pulsar with Iqueye, a novel extremely high time resolution photometer, obtaining data of unprecedented timing accuracy. The data provide the most detailed optical light curve available so far for this pulsar, extending to 27 years the time spanned by X, optical and radio data and allowing a refined determination of the first and second derivatives of the pulsar spin rate.
The resulting value of the braking index, $n$ = 2.087, provides an increasingly consistent evidence that the braking index of the LMC pulsar B0540-69 is slightly higher than $n$ = 2, and definitely smaller than the magnetic dipole value $n$ = 3, in agreement with the findings for all young pulsars for which it has been possible to perform such measurement (with the  possible exception of PSR J1119-6127 for which it has been measured $n$ = 2.91 \citep{cami00}).

\section*{Acknowledgments}

We wish to acknowledge the support given by ESO and the personnel in La Silla Observatory for the perfect operation of Iqueye at NTT.
This work has been supported by the University of Padova, the Italian Ministry of Research and University MIUR, the Italian Institute of Astrophysics INAF, the ÔFondazione Cariparo PadovaÕ, the GALILEO GNSS Supervising Authority through the Harrison Project.
Thanks are due to R. Mignani for the kind help with the finding chart and useful suggestions.
We wish also to thank the referee who pointed out an error in our calculation of the error of the period, and greatly helped to improve  the content.

\label{lastpage}

\end{document}